\documentclass[]{spie}  

\usepackage{amsmath,amsfonts,amssymb}
\usepackage{graphicx}
\usepackage[colorlinks=true, allcolors=blue]{hyperref}

\title{Microfabricated pinholes for high contrast imaging testbeds}

\author[a]{Emory L. Jenkins}
\author[b]{Kyle Van Gorkom}
\author[a]{Kevin Derby}
\author[b]{Patrick Ingraham}
\author[b]{Ewan S. Douglas}
\affil[a]{James C. Wyant College of Optical Sciences, University of Arizona, 1630 E University Blvd, Tucson, AZ 85721, USA}
\affil[b]{Department of Astronomy and Steward Observatory, University of Arizona, 933N Cherry Avenue, Tucson, Arizona, 85721, USA}

\authorinfo{Further author information: (Send correspondence to E.S.D)\\E.S.D.: E-mail: douglase@arizona.edu}

\begin{document} 
\maketitle

\begin{abstract}
In order to reach contrast ratios of $10^{-8}$ and beyond, coronagraph testbeds need source optics that reliably emulate nearly-point-like starlight, with microfabricated pinholes being a compelling solution. To verify, a physical optics model of the Space Coronagraph Optical Bench (SCoOB) source optics, including a finite-difference time-domain (FDTD) pinhole simulation, was created. The results of the FDTD simulation show waveguide-like behavior of pinholes. We designed and fabricated microfabricated pinholes for SCoOB made from an aluminum overcoated silicon nitride film overhanging a silicon wafer substrate, and report characterization of the completed pinholes.
\end{abstract}


\section{INTRODUCTION}
\label{sec:intro}  

The direct imaging of extrasolar bodies is one of the most exciting and most challenging efforts in contemporary astronomy. Extrasolar bodies are not themselves luminous and merely reflect a minute fraction of total solar radiation. For a given exoplanet or debris disk, its host star might outshine it by a factor of $10^8$ or more. An exoplanet, depending on its orbital phase angle and orbit inclination, has small but resolvable angular separation from its host star by modern observatories. There is also a variety of techniques for suppressing starlight that show promise. The main challenge is in wavefront control, since even the slighted aberration can lead to intolerable amounts of starlight leaking through and washing out an exoplanetary image. In order to test coronagraph systems under development in the laboratory, a stand-in `star' source, usually a collimated beam from a pinhole or fiber tip, is used to feed the entrance pupil of the coronagraph. The more similar the source is to starlight, the better the coronagraph can be tested.

\begin{figure} [ht]
\begin{center}
\begin{tabular}{c c}
\includegraphics[height=5cm]{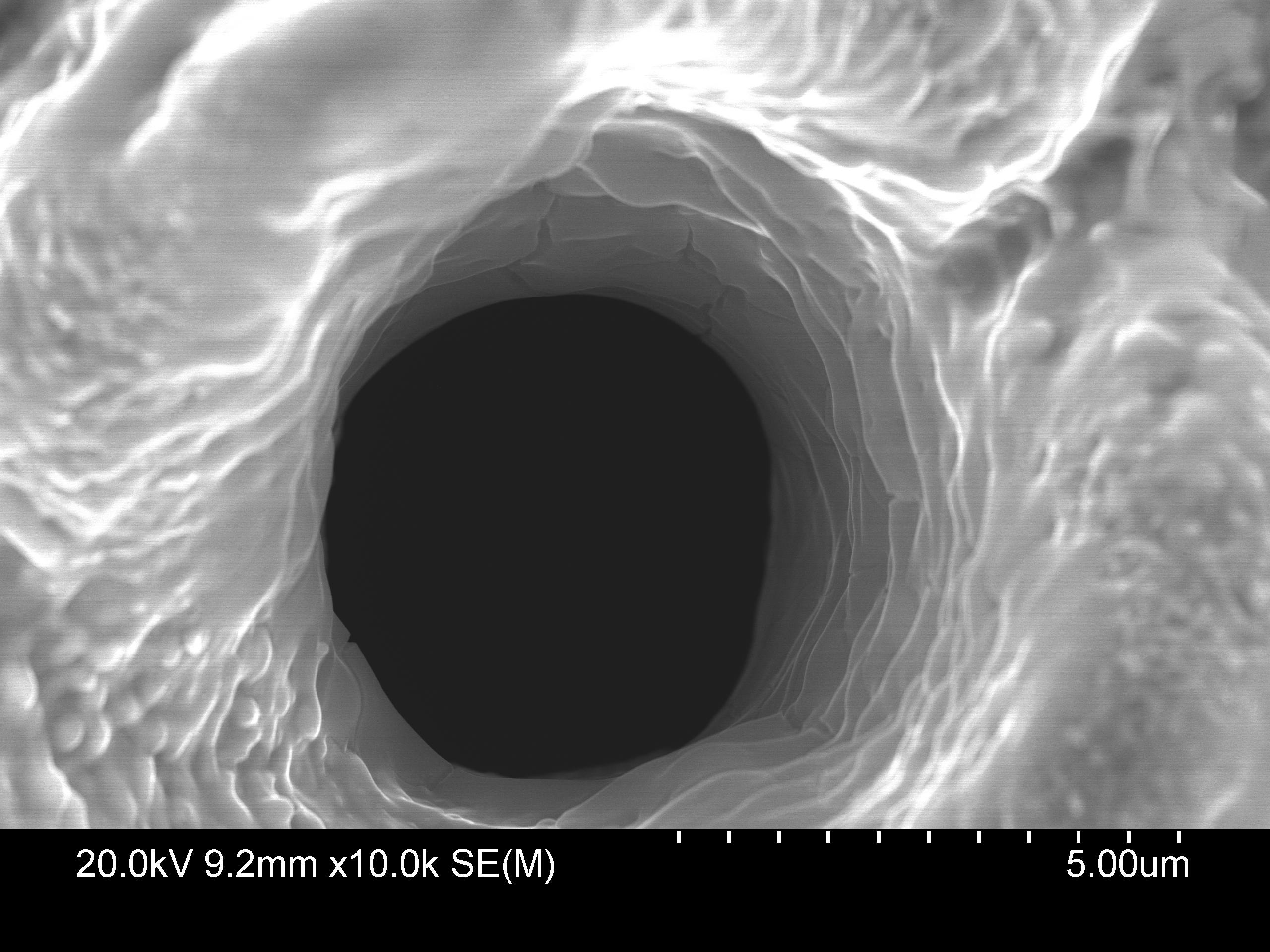}
\end{tabular}
\end{center}
\caption[diagram] 
{ \label{fig:laser_drilled} 
Commercially available $5\pm 1 \mu m$ laser drilled pinhole viewed at two focus depths using a scanning electron microscope at the Kuiper-Arizona Laboratory for Astromaterials Analysis. Note how the pinhole is shaped like a tunnel, with bumps along the sidewalls that can cause back reflections.}
\end{figure}

The University of Arizona Space Astrophysics Lab's (UASAL) Space Coronagraph Optical Bench (SCoOB) testbed \cite{ashcraft2022space,van2022space} is under development as a vacuum-compatible high-contrast imaging and wavefront control testbed \cite{Mawet:09} to test hardware and algorithms for use in future space-based coronagraphs that will image earth-like exoplanets. As the performance of SCoOB improves, there is concern that its source optics will be a limiting factor in achievable contrast. In many situations, it is common to use the small mode field diameter from a single-mode fiber as an approximate point source. However, using a bare single-mode fiber tip is recognized to raise the coronagraphic contrast floor to the $10^{-7}-10^{-8}$ level. Using pinhole spatial filters can improve contrast by increasing beam quality, but off-the-shelf laser-drilled pinholes such as the one shown in Fig. \ref{fig:laser_drilled} have still be known to limit contrast. To address this, UASAL, in collaboration with the University of Arizona Micro/Nano Fabrication Center and the Wyant College of Optical Sciences Micro/Nano Fabrication Cleanroom, have begun microfabricating pinhole spatial filters inspired by those used in the Decadal Survey Testbed's source optics \cite{10.1117/12.2530401}. Our goal is to create higher-quality spatial filters than commercially available laser-drilled pinholes that have previously been used in SCoOB. The design and fabrication of these pinholes is discussed in Sec. \ref{sec:design}.

The size and shape of the pinhole influences the coronagraph's incident wavefront quality, as well as bounding the optical design space. The most basic requirement of the pinhole is that it is smaller than the diffraction limited resolution of the system. For SCoOB, the Airy spot diameter is about $30\mu m$ in object space, so we chose our pinholes to be $\sim 3-8\mu m$ in diameter. At optical wavelengths, a pinhole of this size is $\sim 5\lambda - 20\lambda$ in diameter making the area of the pinhole within one $\lambda$ non-negligible. The interaction with the interface is affected by both the polarization of the electric and magnetic fields and the material properties. The thickness of the pinhole also becomes non-negligible, so the device does not behave as an ideal transmissive mask. Rather, the structure behaves much more similar to a waveguide \cite{Wang:18}, making physical optics propagation unsuitable to model our pinholes. Thus, we used the Finite-Difference Time-Domain (FDTD) method \cite{Goldberg:96} which we discuss in Sec. \ref{subsec:sim}. The testing  of the manufactured pinholes is explained in Sec. \ref{testing}.

\section{Design and Fabrication}
\label{sec:design}
\begin{figure} [ht]
\begin{center}
\begin{tabular}{c}
\includegraphics[height=5cm]{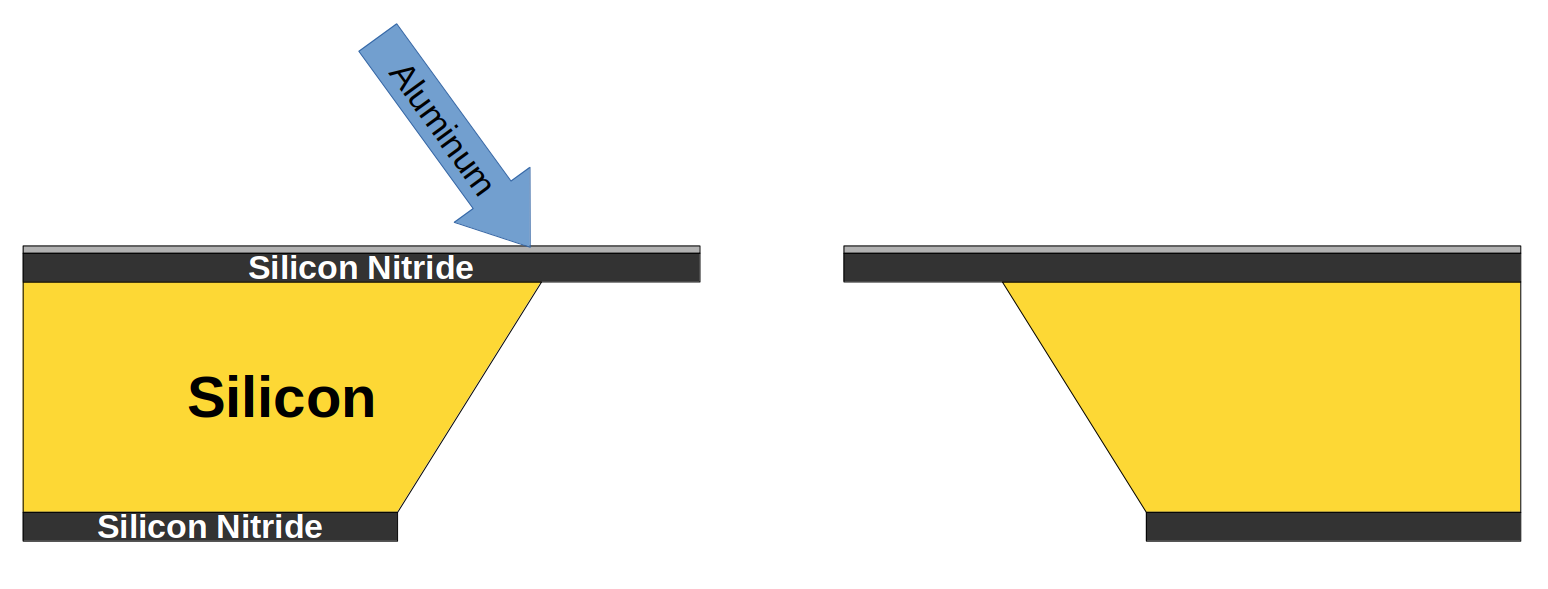}
\end{tabular}
\end{center}
\caption[diagram] 
{ \label{fig:diagram} 
Cross-section of pinhole device. Light propagates from top to bottom.}
\end{figure}

Our pinholes are made in a silicon nitride (Si$_3$N$_4$) membrane that overhangs a silicon wafer substrate as shown in Fig. \ref{fig:diagram}. All patterning was performed by the Wyant College of Optical Sciences Micro/Nano Fabrication Cleanroom and etching and aluminization performed with the assistance of the U of A Micro/Nano Fabrication Center.

Production begins with a $100mm$ $<100>$ orientation $360\mu m$ thick silicon wafer. This wafer has a $1 \mu m$ chemical vapor deposition (CVD) layer of Si$_3$N$_4$ deposited on both surfaces. What will become the rear surface of the wafer is then patterned with negative photoresist using a maskless lithography process. This pattern includes 9 $600\mu m \times 600\mu m$ square windows (one for each pinhole) arranged in a 3 $\times$ 3 grid spaced $25\text{mm}$ apart. Cross-shaped patterns were included at the 4 corners of the wafer far from any pinhole for pattern alignment. Using a reactive ion etcher (RIE), the backside's nitride layer is etched through, exposing the substrate where the square windows and alignment marks are. The front side nitride membrane is then patterned and etched with the nine pinholes measuring $2, 2.5, 3, 3.5, 4, 5, 6, 7, 8 \mu m$ in diameter, one in the center of each window made in the backside nitride layer. The process for patterning and etching the front side was the same as for the back side. We also included 4 fiducials on the front side to help locate the pinholes in testing and alignment. For each pinhole, this consisted of four $100\mu m \times 100\mu m$ windows placed at the edges of a $\pm 1000\mu m$ cross with the pinhole at the center. 

With the front and rear nitride membranes etched, the wafer is then placed in a $30\%$ potassium hydroxide (KOH) wet etch for 6 hours. Silicon nitride is not etched by KOH and forms a barrier such that the silicon substrate is only etched where the nitride membrane has been exposed. The KOH wet etch is an anisotropic process, heavily favoring etching the $<100>$ plane while removing negligible material in the $<111>$ plane. Thus, the KOH etches at $54.7^\circ$ rather than $90^\circ$ to the wafer surface. This results in the square openings on the backside forming the base of a 4-sided pyramidal volume being removed in the silicon substrate up until the pinhole on the front side. The reliefs behind each pinhole are sized such that a roughly $90\mu m \times 90\mu m$ square region around each pinhole overhangs the silicon substrate, and the pinhole indicator marks are sized such that the silicon cannot be etched all the way through.

Since the nitride layer is transmissive in the visible spectrum, after the wet etch a $200nm$ aluminum film is sputtered to the front side of the wafer to achieve opacity. At $632.8 nm$, the intensity transmission through this film should be no more than $10^{-13}$. The wafer is then cleaved to separate the pinholes into individual chips, though since we processed two wafers, one has been kept uncleaved as backup in the event of a failure. The final step is to adhere the pinholes with vacuum-compatible epoxy to an in-house machined aluminum annulus that fits in standard 1-inch optical mounts.

\section{Testing of fabricated wafers/pinholes}
\label{testing}

\begin{figure} [ht]
\begin{center}
\includegraphics[height=12cm]{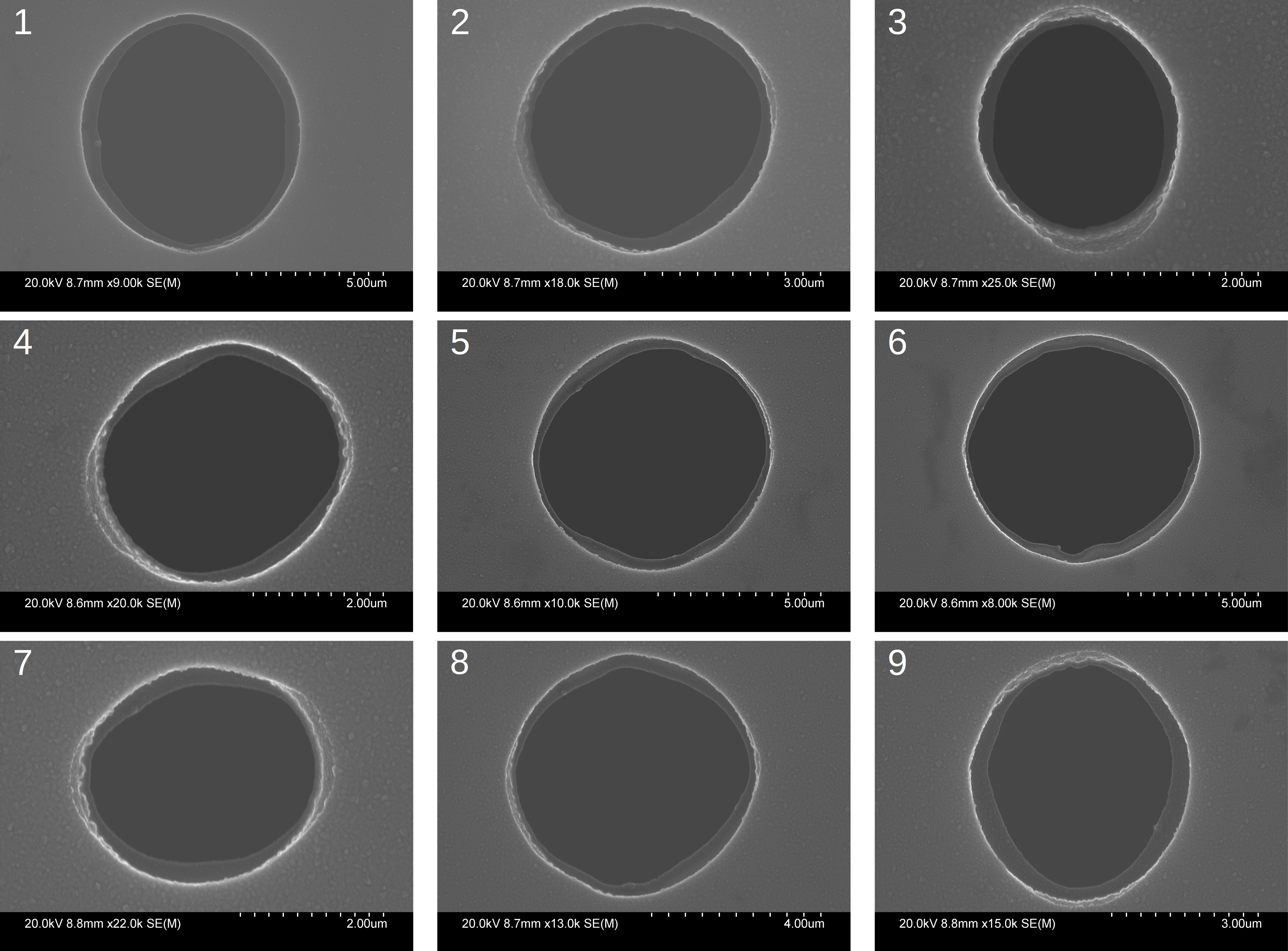}
\end{center}
\caption[example] 
{ \label{fig:SEM} 
Nine pinholes from the cleaved wafer under various magnifications with an SEM with the aluminized side facing towards viewer. Some sections of the front edges have rough features, most often along the elongation.}
\end{figure}

The wafer that was cleaved yielded all 9 pinholes and each were examined using a scanning electron microscope (SEM), shown in Fig. \ref{fig:SEM}. Indeed, all are to an extent elliptical, thought the departure from circular is more extreme for the smaller pinholes. This indicates that we are nearing the limits of our patterning and etching processes. Another feature of note is that some portions of the pinhole edge are significantly rougher, particularly the areas located at the ends of the major axis if an ellipse were to be fitted, whereas elsewhere the edge is quite sharp. The rear pinhole edge is visible from this `top-down' perspective, which is expected due to the RIE creating $\sim 85^\circ$ sidewalls. It is clear, however, that at these scales the sidewall angle has significant variance. Pinhole 6 shows the strongest deviation local deviation with what resembles a shark tooth in its rear edge.

\begin{figure} [ht]
\begin{center}
\begin{tabular}{c c} 
\includegraphics[height=6cm]{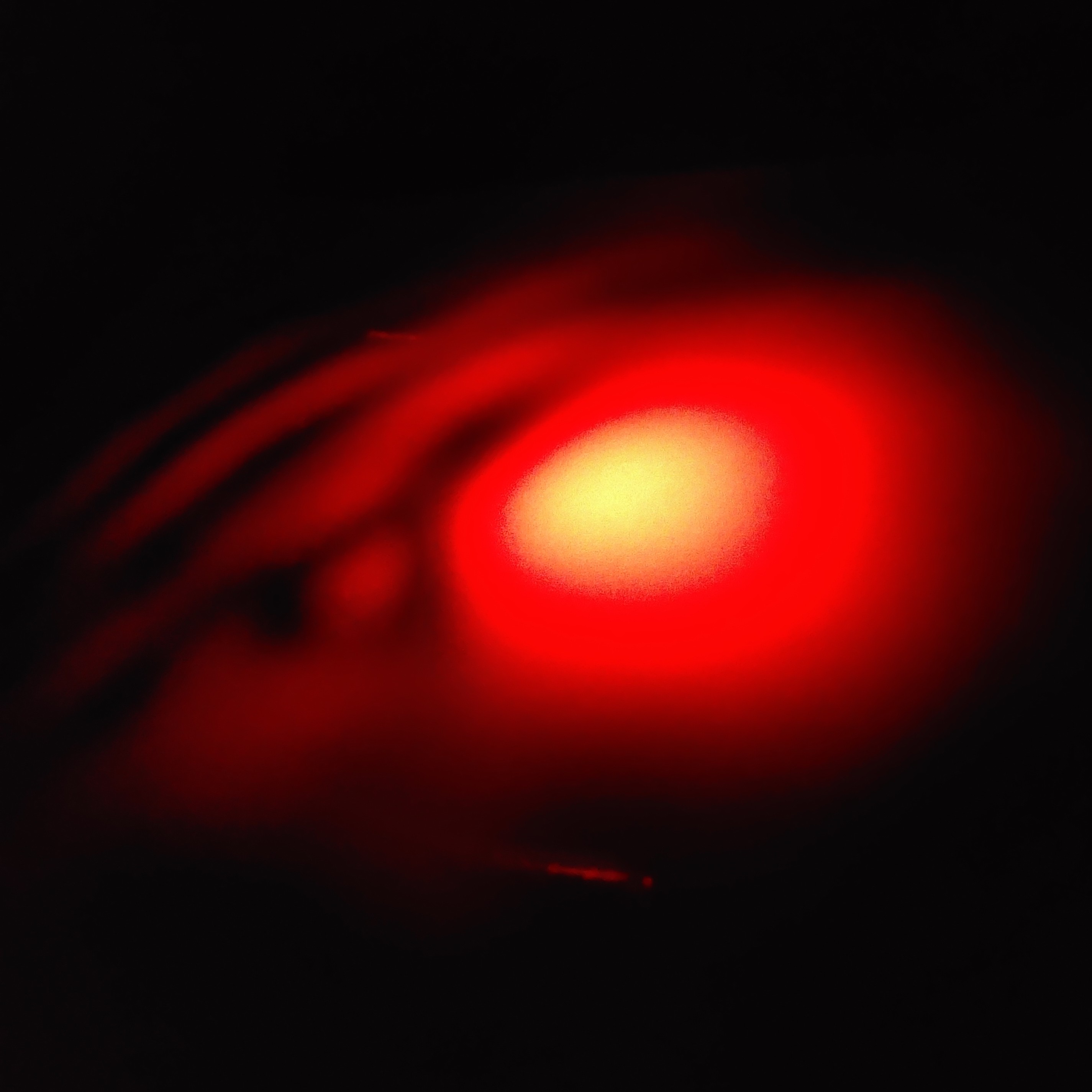} &
\includegraphics[height=6cm]{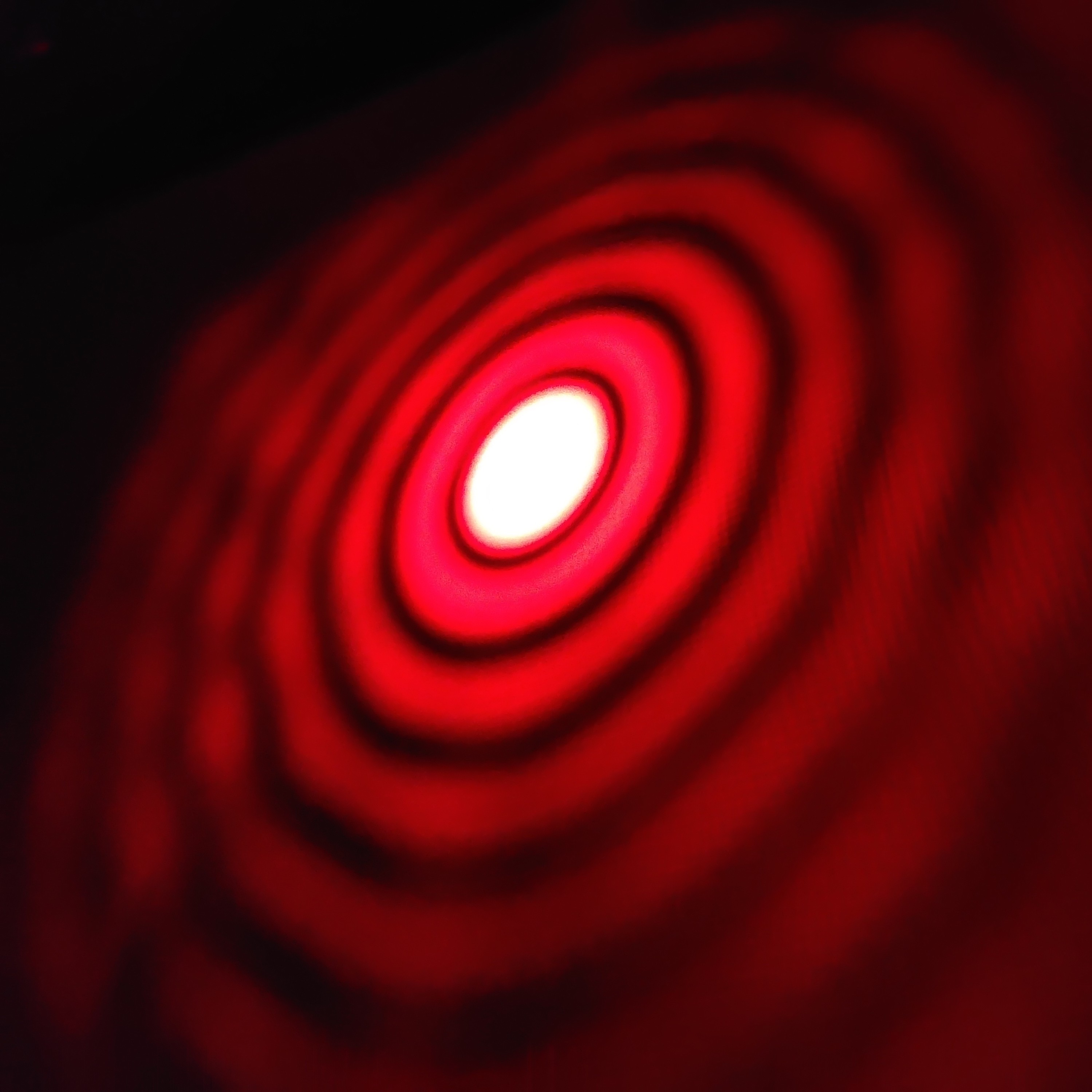} 
\end{tabular}
\end{center}
\caption[example] 
{ \label{fig:diffraction} 
Diffraction pattern from a commercial laser-drilled pinhole (left) and a microfabricated pinhole (right). The large number of rings from the microfabricated pinhole indicates a large improvement in beam quality.}
\end{figure}

As a qualitative assessment of the quality of the microfabricated pinholes, pinhole 6 and the laser-drilled pinhole from Fig. \ref{fig:laser_drilled} were illuminated with a helium-neon (HeNe) laser and the diffraction patterns were compared. For the laser-drilled pinhole, we were unable to clearly decipher the first diffraction ring of the Airy pattern. Pinhole 6, despite having a sharp feature on the edge, showed at least 6 rings, indicating a much higher beam quality. One unexpected feature of the diffracted beam from the microfabricated pinhole is a high spatial frequency fringe pattern, visible in the right portion of the image in Fig. \ref{fig:diffraction}. In the lab, this pattern was noticeable throughout the Airy pattern, while these features were out of focus in much of the image and less apparent. One hypothesis for this phenomenon is reflections from high diffraction angle light on the pyramidal relief in the bulk silicon. It's not yet known whether this effect is detrimental on coronagraph performance. If this is the case, a solution may be to aluminize the rear side of the wafer and run the pinhole in ``reverse" configuration.

\subsection{Simulation}
\label{subsec:sim}
In order to create a comprehensive model of the pinhole used to illuminate the SCoOB testbed and vector-vortex coronagraphic (VVC) mask, our simulation is broken down into 4 subroutines:
\begin{enumerate}
    \item Pre-pinhole fiber source relay
    \item FDTD intra-pinhole propagation
    \item Pinhole to pupil relay
    \item VVC and contrast calculation
\end{enumerate}

The first routine begins with a `fiber tip' represented by a $4\mu m$ full-width half-max (FWHM) gaussian amplitude profile with flat phase at wavelength of the HeNe laser ($632.8 nm$). This is Fraunhofer propagated to a thin lens that collimates the beam. Fresnel propagation via the Physical Optics Propagation in PYthon (POPPY) \cite{perrin2012poppy} package to a second lens followed by a final Fraunhofer propagation relays the fiber tip to an image plane where the pinhole is located with unit magnification.

The FDTD method is a numerical solution to Maxwell's equations and thus captures the full electric and magnetic fields. For this reason, it is often used to model EM fields interacting with structures sized on the order of the wavelength, where boundary conditions may result in fields destructively interfering or not depending on the medium of incidence. Since our pinholes have a non-negligible thickness, they behave as a short waveguide, with light reflecting off the inner surface rather than being absorbed completely.

For our FDTD simulation, we use the Meep (MIT Electromagnetic Equation Propagation) \cite{oskooi2010meep} python package. A $10 \times 10 \times 3 \mu m$ spatial grid is used with a resolution of 80 voxels per micron. The structure is set up with a $10 \times 10 \times 1 \mu m$ silicon slab attached to a $10 \times 10 \times 0.2 \mu m$ aluminum slab centered in the simulation grid. The pinhole is created by using a custom material function to inscribe an extruded ellipse within the slabs and set the material to vacuum. This ellipse is adjustable in size, flatness, and major axis angle so that our simulated pinholes can match our microfabricated pinholes in shape. Due to the plasma etching of nitride generally creating side walls at $\sim 85^\circ$ relative to the wafer, an adjustable cone angle was added to the material function. While Meep is distributed with a materials library that includes Si3N4 and Al, custom dispersion free pseudo-materials are initialized using the conductivities and permittivities of Si3N4 and Al at the source wavelength. The advantage is a sizeable improvement in memory consumption as well as computation time.

The source is set up as a ContinuousSource plane source located $0.2mm$ before the aluminum surface to allow light to propagate just a small amount before reaching the pinhole. The relayed Gaussian wavefront from the fiber is loaded from a pickle file into the ContinuousSource object as a complex amplitude profile ndarray and interpolated by Meep to the simulation grid as either $x$ or $y$ polarization. The simulation runs for $\sim 13fs$ which is enough time to settle into steady state and corresponds to about $6$ optical cycles at $632.8 nm$. To capture the effects of feeding a pinhole unpolarized light, the FDTD simulation is performed twice, once with $x$ and once with y-polarized light.

The FDTD simulation outputs non-zero field in x, y, and z polarizations for inputs of purely $x$ or y. Though we are not interested in propagation of the z-polarized light, the two FDTD simulations, one for $x$ and one for $y$ input polarizations, each result in a pair of correlated $x$ and $y$ linearly polarized wavefronts. A matrix Fourier transform (MFT) is used to propagate each of these wavefronts to a thin collimating lens in order to maintain appropriate sampling in the pupil plane from such a small object size without using unreasonably large arrays or downsampling \cite{MFT}. This is followed by a Fresnel propagation in POPPY to relay the beam into the entrance pupil of a simple VVC model.

VVC modelling is done using the High Contrast Imaging for Python (HCIPy) \cite{por2018hcipy} package which has native support for polarized wavefronts. The model consists of a linear polarizer with $10,000:1$ extinction followed by a quarter waveplate (QWP) with $0.2515\lambda$ retardance to circularize the incoming beam in the pupil plane. This extiction ratio and retardance were chosen to match the performance of the optics in SCoOB. The wavefront is then propagated to the Lyot stop plane using HCIPy's built-in VVC  with the vector vortex waveplate (VVW) set to $178^\circ$ retardance and charge-6, matching SCoOB. The Lyot stop is $90\%$ the diameter of the entrance pupil and rejects the remaining `starlight.' Another QWP and linear polarizer with the same specs as before are used to select the circular polarization previously generated, and the now linearly polarized exit pupil is propagated to an image plane. No surface errors for any of the polarization optics are included in the model.

The VVC model is run twice, once for entrance entrance pupil field corresponding to the $x$ and $y$ polarization pinhole inputs, and the two resulting PSFs are incoherently summed to give the full unpolarized pinhole illumination PSF. For contrast measurements, the same entrance pupils are propagated to the image plane without the polarization optics or VVC inserted in the model and the PSFs are summed in intensity to give a reference PSF. The maximum intensity of these reference PSFs are used to normalize the VVC PSFs to convert to units of contrast.

\subsection{Simulation Results}
\begin{figure} [ht]
\begin{center}
\begin{tabular}{c c} 
\includegraphics[height=5cm]{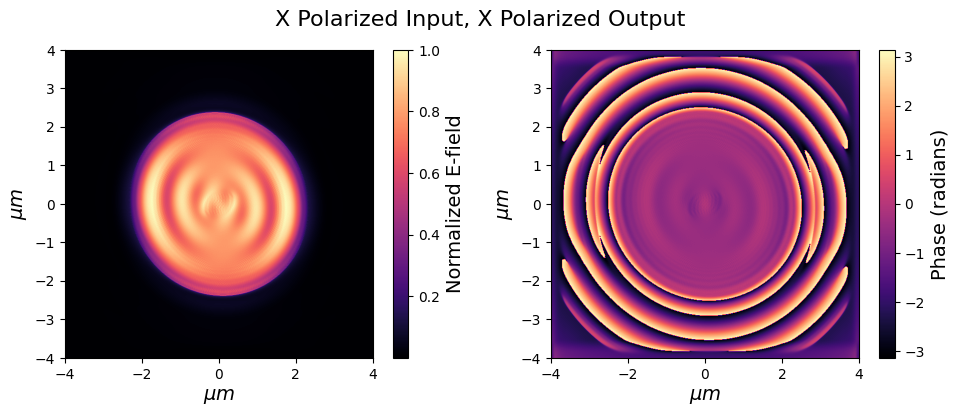} \\
\includegraphics[height=5cm]{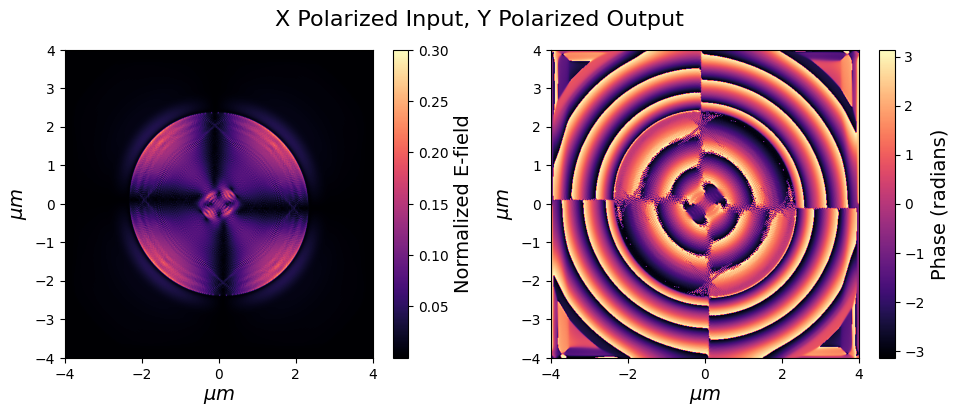} \\
\includegraphics[height=5cm]{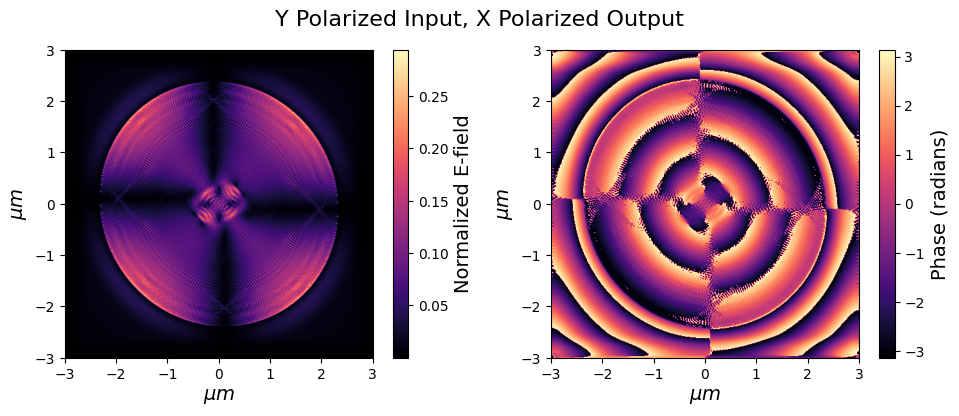} \\
\includegraphics[height=5cm]{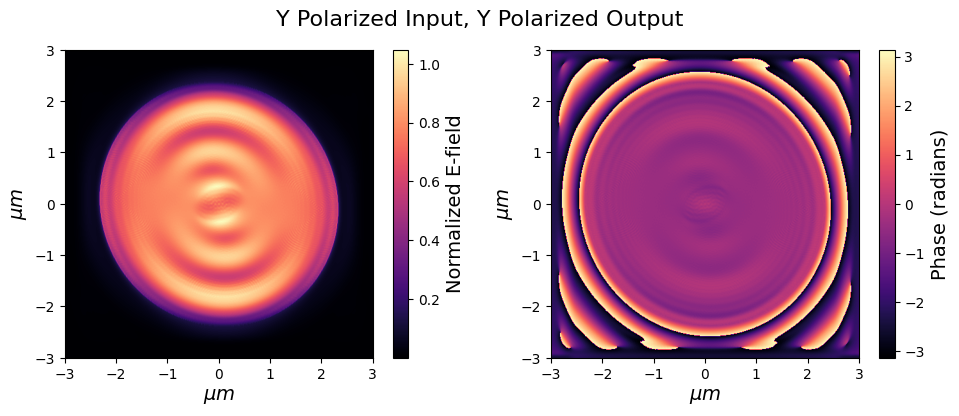} 
\end{tabular}
\end{center}
\caption[example] 
{ \label{fig:pinhole_out} 
Top: magnitude and phase of $x$ polarized (left) and $y$ polarized (right) electric fields at rear of a pinhole for $x$ polarized input. Bottom: magnitude and phase of $x$ polarized (left) and $y$ polarized (right) electric fields at rear of a pinhole for $y$ polarized input. This demonstrates the significant coupling between polarization states in the pinhole.}
\end{figure}

The FDTD simulation shows that there is significant coupling between polarization states. Take for example a $5\mu m$ pinhole with $5\%$ flatness, $30^\circ$ fast axis orientation, and $85^\circ$ sidewalls, shown in Figure \ref{fig:pinhole_out}. The presence of $y$ polarized electric field in the output for a purely $x$ polarized input and vice-versa, along with the spatial non-uniformity of the output field corroborates our assumption that the thickness of the pinhole is non-negligible. Generating PSFs with the resulting wavefronts from the pinholes using a coronagraph model is yielding promising results which will be featured in an upcoming publication.

\section{Conclusions and Future Work}
We developed microfabricated precision pinholes for use as pseudo-point sources in high contrast imaging testbeds as star substitutes. The diffraction patterns they create are much higher quality than the previously used laser-drilled pinholes, making them a promising solution for SCoOB's source. We are simulating the performance of a VVC with these microfabricated pinholes and will provide contrast results in a future publication. Implementation of a pinhole on the testbed is in progress, and expected to be completed by the end of 2023. In addition to comparing dark holes dug with a microfabricated pinhole to those dug with a laser-drilled one on SCOoB, the diffracted light from the microfabricated pinholes will be studied with a polarimeter. Finally, we plan to investigate the impact of the high spatial frequency patterns seen in the diffracted beam and will test mitigation strategies if necessary.

\acknowledgments 
Portions of this research were supported by funding from the Technology Research Initiative Fund (TRIF) of the Arizona Board of Regents
and by generous anonymous philanthropic donations to the Steward Observatory of the College of Science at the University of Arizona.
This work was performed in part at the Micro/Nano Fabrication Center at the University of Arizona
We acknowledge NASA grants \#NNX12AL47G, \#NNX15AJ22G and \#NNX07AI520, and NSF grants \#1531243 and \#EAR-0841669 for funding of the instrumentation in the Kuiper-Arizona Laboratory for Astromaterials Analysis at the University of Arizona.
The authors would like to thank James Bohlman and Dr Greg Book from the U of A Micro/Nano Fabrication Center as well as Roland Himmelhuber and Joe Chamberlain from the Wyant College of Optical Sciences Micro/Nano Fabrication Cleanroom for their support in developing a process and creating these pinholes. Thanks to Dr Jerry Chang from the Kuiper-Arizona Laboratory for Astromaterials Analysis for SEM training.

\bibliography{report} 
\bibliographystyle{spiebib} 

\end{document}